\pdfoutput=1
\documentclass[twocolumn]{aastex631}

\hypersetup{
  colorlinks=true,
  linkcolor=blue,
  citecolor=blue,
  urlcolor=blue,
  filecolor=blue
}

\usepackage{amsmath,amssymb,mathtools,comment}

\makeatletter
\@ifundefined{tablenum}{}{}
\makeatother
\usepackage{siunitx}
\shorttitle{Time-Delay Asymmetry in Kerr}
\shortauthors{Chowdhury}

\begin{document}

\title{A New Timing Signature of Black Hole Spin: Time-Delay Asymmetry in Kerr Accretion Flows}

\correspondingauthor{Shakibul Chowdhury}
\email{shakibul.chowdhury@gmail.com}

\author{Shakibul Chowdhury}

\affiliation{Department of Physics, The City College of New York, New York, NY 10031, USA}

\begin{abstract}

We introduce a new general-relativistic timing observable that measures the breaking of reflection symmetry in photon arrival times caused by black hole spin. Using backward ray tracing in the Kerr spacetime, we construct time-delay maps across the observer image plane and define a mirror-paired asymmetry based on photons arriving from opposite sides of the projected spin axis. In the Schwarzschild limit ($a=0$), the asymmetry vanishes to numerical precision, providing a stringent validation test of the method. For rotating black holes, Kerr rotation breaks the left-right propagation symmetry of null geodesics, producing systematic differences between prograde and retrograde photon trajectories and resulting in a nonzero mirror-paired timing asymmetry, $A_t$. We find that $A_t$ increases with spin and depends strongly on observer inclination and emission radius, with the largest signals arising from emission close to the black hole and from intermediate to high inclinations. Converting the dimensionless asymmetry into physical units yields timing offsets ranging from seconds to hours for representative supermassive black hole systems. Unlike traditional timing analyses based on spatially integrated signals, the observable introduced here isolates directional information encoded in Kerr photon propagation and provides a physically motivated timing signature of black hole rotation. We discuss the implications of this effect for strong-gravity timing studies and X-ray reverberation mapping.

\end{abstract}


% ---------- 1. Introduction ----------
\section{Introduction}\label{sec:intro}

Photon propagation in the curved spacetime surrounding black holes produces a wide range of observable relativistic signatures, including gravitational lensing, Doppler boosting, redshift effects, polarization rotation, and time delays \citep{Cunningham1975,Laor1991,SchnittmanKrolik2010}. These phenomena provide powerful probes of strong-field gravity and are routinely used to infer the properties of accreting black hole systems through spectral, imaging, polarimetric, and timing observations \citep{Reynolds2021}. Recent horizon-scale imaging observations have further highlighted the role of relativistic photon propagation, gravitational lensing, and asymmetric image morphology near Kerr black holes \citep{EHT2019_V,EHT2019_VI,Gralla2019}. In particular, X-ray timing techniques such as reverberation mapping have emerged as powerful probes of the geometry and dynamics of accretion flows near compact objects \citep{Reynolds1999,WilkinsFabian2013,Uttley2014,Cackett2021}.

A central goal of strong-gravity astrophysics is the measurement of black hole spin. In Kerr spacetime, black hole rotation breaks the symmetry between prograde and retrograde photon propagation paths, producing directional asymmetries in null geodesic trajectories \citep{Bardeen1972,Fanton1997,BeckwithDone2005}. Observational constraints on spin are commonly obtained through relativistic reflection spectroscopy, iron-line modeling, continuum fitting, and reverberation analyses \citep{Miller2007,Reynolds2021}. However, many existing timing observables are based on spatially integrated signals and therefore average over directional information encoded in photon propagation across the observer image plane.

Time delays provide a particularly direct probe of relativistic photon propagation. The arrival time of a photon depends not only on the geometry of the emitting region, but also on the structure of the underlying spacetime metric through which the photon travels. While relativistic transfer functions and reverberation delays have been studied extensively \citep{Cunningham1975,WilkinsFabian2012,Emmanoulopoulos2014,Uttley2014,Cackett2021}, comparatively little attention has been given to the symmetry properties of arrival-time distributions across the image plane itself.

In this work, we introduce a new timing observable that quantifies the breaking of reflection symmetry in photon arrival times induced by black hole spin. Using general-relativistic ray tracing in the Kerr metric, we construct time-delay maps across the observer image plane and compare photon trajectories related by reflection across the projected spin axis. From these mirror-paired trajectories, we define a scalar timing asymmetry that vanishes in the Schwarzschild limit and becomes nonzero in rotating spacetimes. Physically, the asymmetry arises because Kerr rotation modifies null geodesic propagation differently on opposite sides of the projected spin axis.

Our approach isolates a geometric and propagation-based signature of spin that is independent of detailed radiative transfer assumptions. This makes the observable conceptually simple and potentially robust against uncertainties in emission physics and accretion-flow structure. By systematically exploring its dependence on spin, inclination, and emission radius, we show that the asymmetry provides a systematic and physically interpretable timing signature of Kerr spacetime.

The structure of this paper is as follows. In Section~\ref{sec:theory}, we summarize the theoretical framework for photon propagation in the Kerr metric. In Section~\ref{sec:methods}, we describe the numerical ray-tracing framework used to construct time-delay maps. In Section~\ref{sec:asymmetry}, we define the mirror-paired timing asymmetry observable. Section~\ref{sec:results} presents the main numerical results and parameter dependencies. In Section~\ref{sec:discussion}, we discuss observational implications and astrophysical scaling, and we conclude in Section~\ref{sec:conclusion}.

% ---------- 2. Theoretical Framework ----------
\section{Theoretical Framework}\label{sec:theory}

\subsection{Kerr Spacetime}

The spacetime surrounding a rotating black hole is described by the Kerr metric, which represents the unique stationary, axisymmetric vacuum solution of Einstein’s field equations characterized by mass $M$ and angular momentum $J$ \citep{Kerr1963,Bardeen1972}. In Boyer--Lindquist coordinates $(t,r,\theta,\phi)$, the Kerr line element is given by

\begin{align}
ds^2 =\,
&-\left(1-\frac{2Mr}{\Sigma}\right)dt^2
-\frac{4Mar\sin^2\theta}{\Sigma}\,dt\,d\phi
\nonumber\\[4pt]
&+\frac{\Sigma}{\Delta}dr^2
+\Sigma\,d\theta^2
\nonumber\\[4pt]
&+\left(
r^2+a^2+
\frac{2Ma^2r\sin^2\theta}{\Sigma}
\right)
\sin^2\theta\,d\phi^2 .
\end{align}

where

\begin{equation}
\Sigma = r^2 + a^2\cos^2\theta,
\end{equation}

and

\begin{equation}
\Delta = r^2 - 2Mr + a^2.
\end{equation}

Here,

\begin{equation}
a = \frac{J}{M}
\end{equation}

is the spin parameter of the black hole. Throughout this work, we adopt geometrized units with

\begin{equation}
G = c = 1.
\end{equation}

In the limit $a=0$, the metric reduces to the Schwarzschild spacetime, which is spherically symmetric.

\subsection{Photon Propagation and Frame Dragging}

Photon trajectories in Kerr spacetime follow null geodesics satisfying

\begin{equation}
ds^2 = 0.
\end{equation}

The curvature of spacetime modifies both the spatial trajectories and arrival times of photons propagating near the black hole \citep{Cunningham1975,Laor1991,BeckwithDone2005}. In rotating spacetimes, frame dragging introduces an additional asymmetry between photons moving in the prograde and retrograde directions relative to the black hole spin \citep{Bardeen1972}.

This effect originates from the off-diagonal metric term

\begin{equation}
g_{t\phi},
\end{equation}

which couples temporal and azimuthal motion. As a consequence, photons traveling on opposite sides of the projected spin axis experience different effective path lengths and coordinate propagation times.

The resulting asymmetry is a geometric consequence of frame dragging and relativistic photon propagation in Kerr spacetime.

\subsection{Observer Image Plane}

To characterize photon arrival times, we construct an observer image plane parameterized by Cartesian coordinates $(\alpha,\beta)$ corresponding to impact parameters on the sky \citep{Cunningham1975,Fanton1997,Vincent2011}. Each point on the image plane defines an initial null geodesic launched backward from a distant observer toward the black hole.

For a given trajectory, the coordinate arrival-time delay is defined as

\begin{equation}
t(\alpha,\beta)
=
t_{\rm arrival}
-
t_{\rm emission},
\end{equation}

where $t_{\rm emission}$ is the coordinate time at which the photon intersects the emitting region and $t_{\rm arrival}$ is the corresponding coordinate arrival time at the observer.

The collection of these delays defines a two-dimensional time-delay map over the observer image plane.

\subsection{Reflection Symmetry}

In Schwarzschild spacetime, the spherical symmetry of the metric implies that photon trajectories related by reflection across the image-plane symmetry axis satisfy

\begin{equation}
t(\alpha,\beta)=t(-\alpha,\beta),
\end{equation}

up to numerical precision. Consequently, any mirror-paired timing asymmetry vanishes in the nonrotating limit.

In Kerr spacetime, however, frame dragging breaks this reflection symmetry. Prograde and retrograde photon trajectories experience different propagation times, producing a systematic asymmetry in the arrival-time distribution across the image plane.

The timing observable introduced in this work is designed specifically to quantify this symmetry breaking.

% ---------- 3. Numerical Methods ----------
\section{Numerical Methods}\label{sec:methods}

\subsection{Ray-Tracing Framework}

Photon trajectories are computed using backward ray tracing from a distant observer to the emitting region in Kerr spacetime. For each pixel on the observer image plane, labeled by coordinates $(\alpha,\beta)$, we construct an initial null four-momentum and integrate the geodesic equations backward in affine parameter.

The geodesic equations are evolved using a fourth-order Runge--Kutta (RK4) integration scheme with the Kerr metric expressed in Boyer--Lindquist coordinates. Photon trajectories are followed until they either intersect the emitting surface, cross the event horizon, or fail to intersect the emitting region.

Our implementation follows standard general relativistic ray-tracing approaches commonly employed in studies of black hole imaging, relativistic transfer functions, and polarization transport \citep{Cunningham1975,Laor1991,Fanton1997,SchnittmanKrolik2010,Vincent2011}. Elements of the numerical integration framework and observer-plane construction build upon the relativistic transport infrastructure developed in previous relativistic polarization transport work \citet{Chowdhury2026}.

\subsection{Emission Geometry}

We model the emitting region as a geometrically thin accretion disk confined to the equatorial plane,

\begin{equation}
\theta = \frac{\pi}{2}.
\end{equation}

Photon trajectories are evaluated at fixed emission radii $R_{\rm emit}$ in order to isolate the geometric effects of relativistic photon propagation in Kerr spacetime. Only photon trajectories that intersect the emitting surface are retained for further analysis.

At the intersection point, the coordinate emission time $t_{\rm emission}$ is recorded.

\subsection{Image-Plane Sampling}

The observer image plane is discretized using a uniform Cartesian grid in $(\alpha,\beta)$. A null geodesic is computed independently for each grid point.

Only pixels corresponding to rays that successfully intersect the emitting surface are retained. Rays that fail to intersect the emission region or cross the event horizon are excluded from the timing analysis. This ensures that the resulting calculations contain only physically relevant photon trajectories.

Mirror-paired pixels reflected across the image-plane symmetry axis are subsequently used to construct the timing-asymmetry observable introduced in Section~\ref{sec:asymmetry}.

\subsection{Parameter Space}

We compute photon trajectories over a range of black hole spins, observer inclinations, and emission radii. The dimensionless spin parameter range considered in this work is

\begin{equation}
0 \le a/M \le 0.998,
\end{equation}

thereby probing both weakly rotating and near-extremal Kerr spacetimes.

Observer inclinations of
\[
i = 10^\circ,\ 30^\circ,\ 50^\circ,\ 75^\circ
\]
are used to probe viewing geometries ranging from near face-on to highly inclined configurations.

Emission radii are selected to probe both strong-field and moderately weak-field regimes. Representative emission radii adopted in this work are

\begin{equation}
R_{\rm emit} = 4.5M,\ 5.2M,\ 6.0M.
\end{equation}

Representative visualizations and scaling examples shown throughout the paper frequently adopt the intermediate inclination $i=50^\circ$, which provides a clear illustration of the spin-induced timing asymmetry while avoiding the limiting near-face-on and extreme edge-on viewing geometries.

\subsection{Numerical Validation and Convergence}

The accuracy of the geodesic integration is verified by monitoring the null normalization condition of the photon four-momentum throughout each trajectory, following standard practices in relativistic ray tracing.

Additional validation is provided by the Schwarzschild limit ($a=0$), where the expected reflection symmetry of photon propagation is recovered to numerical precision.

Additional convergence tests with respect to image-plane resolution and affine-parameter step size are presented in Appendix~\ref{sec:appendixB}.

The measured timing asymmetry remains stable under increasing numerical refinement, demonstrating that the observed signal arises from relativistic photon propagation rather than numerical artifacts.

% ---------- 4. Time-Delay Maps ----------
\section{Time-Delay Maps}\label{sec:maps}

The primary output of the ray-tracing framework is a two-dimensional map of photon arrival-time delays across the observer image plane. The resulting function $t(\alpha,\beta)$ encodes the relativistic propagation structure of Kerr spacetime.

Variations in arrival time arise from differences in photon path length, gravitational time dilation, and frame-dragging effects near the rotating black hole \citep{Cunningham1975,WilkinsFabian2012,Uttley2014}.

In the Schwarzschild limit, the delay structure remains reflection symmetric about the image-plane symmetry axis. In contrast, Kerr rotation breaks the reflection symmetry of null-geodesic propagation across the observer image plane, producing asymmetric arrival-time structure between prograde and retrograde photon trajectories \citep{Bardeen1972,BeckwithDone2005}.

\begin{figure*}
    \centering
    \includegraphics[width=0.95\textwidth]{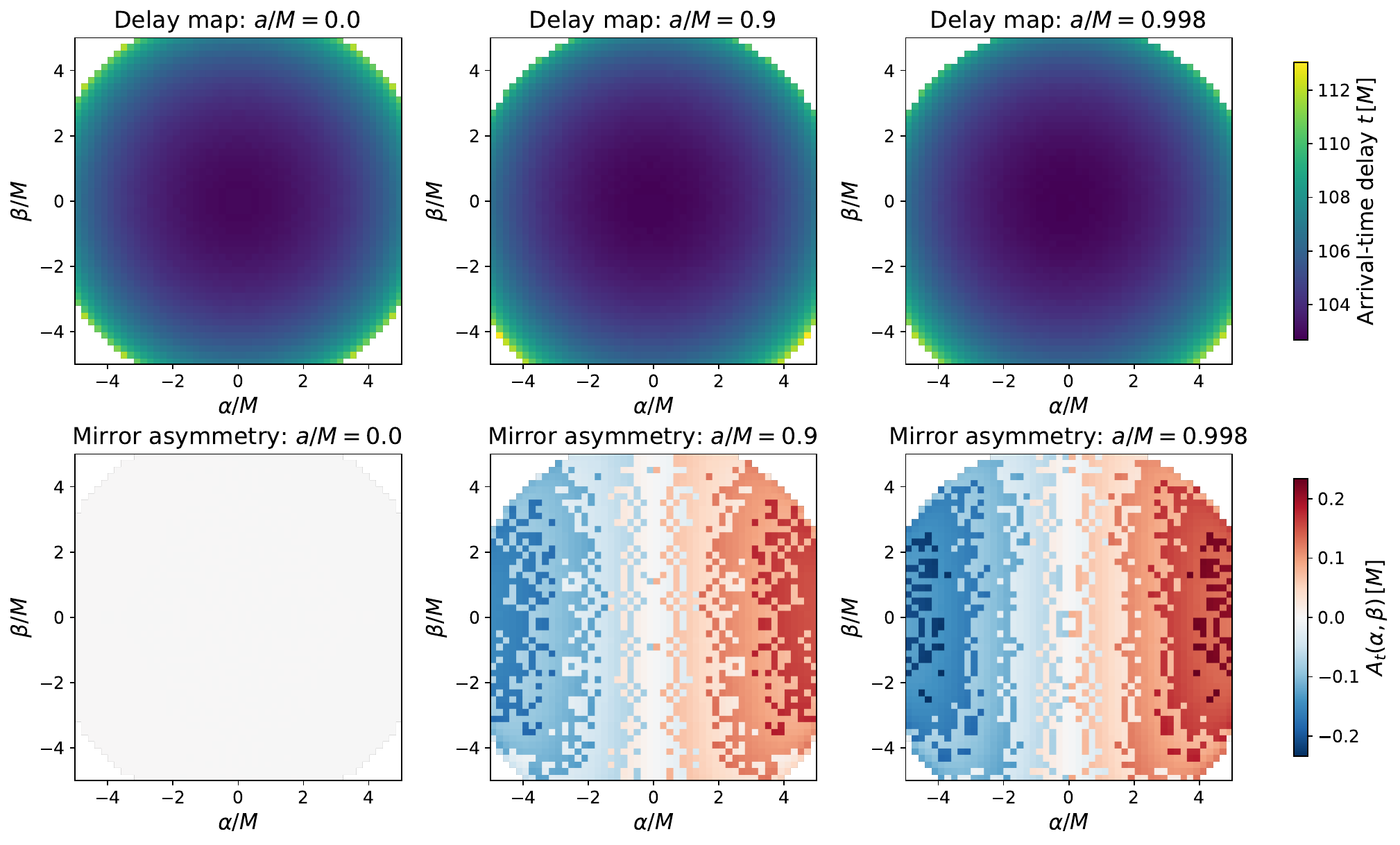}
    \caption{
    Representative arrival-time delay maps and mirror-paired asymmetry maps for 
    $a/M=0$, $0.9$, and $0.998$, shown for a fixed observer inclination and emission radius.
    The top row shows the coordinate arrival-time delay $t(\alpha,\beta)$ across the observer image plane.
    The bottom row shows the corresponding mirror asymmetry 
    $\Delta t(\alpha,\beta)=t(-\alpha,\beta)-t(\alpha,\beta)$.
    In the Schwarzschild limit, the asymmetry vanishes to numerical precision, confirming the expected reflection symmetry.
    For rotating Kerr black holes, the asymmetry becomes nonzero and develops a coherent antisymmetric structure across the image plane, reflecting the differential propagation times of prograde and retrograde photon trajectories induced by frame dragging.
    }
    \label{fig:delay_maps}
\end{figure*}

Figure~\ref{fig:delay_maps} presents representative photon arrival-time delay maps and the corresponding mirror-paired asymmetry maps for increasing black hole spin. The top row demonstrates that the delay structure remains smooth across the observer image plane, while the bottom row isolates the parity-breaking timing structure induced by Kerr rotation.

In the Schwarzschild case, the mirror asymmetry vanishes to numerical precision, consistent with the underlying reflection symmetry of the spacetime. For rotating Kerr black holes, however, the asymmetry becomes systematically nonzero and develops a coherent antisymmetric structure across the image plane. These directional asymmetries demonstrate that frame dragging modifies the propagation times of photons arriving from opposite sides of the observer image plane and motivate the mirror-paired timing observable introduced in the following section.

\section{Time-Delay Asymmetry}\label{sec:asymmetry}

We now construct a quantitative timing observable directly from the delay maps introduced in Section~\ref{sec:maps}.

Unlike traditional timing analyses that focus on spatially integrated transfer functions or temporal variability, our approach preserves directional information across the observer image plane. Because the asymmetry may change sign across the image plane, spatial integration can partially suppress the signal through cancellation.

The mirror-paired construction introduced here instead isolates parity-breaking timing structure associated with relativistic photon propagation in Kerr spacetime \citep{Cunningham1975,WilkinsFabian2012,Uttley2014}.

\subsection{Mirror Pairing}
To quantify reflection-symmetry breaking, we consider mirror-paired pixels across the projected spin axis. For a given point $(\alpha,\beta)$, the corresponding reflected point is defined by

\begin{equation}
(\alpha,\beta)
\rightarrow
(-\alpha,\beta).
\end{equation}

We then define the mirror-paired timing difference

\begin{equation}
\Delta t(\alpha,\beta)
=
t(-\alpha,\beta)
-
t(\alpha,\beta),
\end{equation}

evaluated only for pairs where both trajectories intersect the emitting surface.

This construction isolates directional timing differences between null geodesics propagating on opposite sides of Kerr spacetime.

\subsection{Asymmetry Observable}

To characterize the overall magnitude of the symmetry breaking, we define the root-mean-square timing asymmetry

\begin{equation}
A_{t,\rm rms}
=
\left<
\Delta t^2
\right>^{1/2},
\end{equation}

where the average is taken over all valid mirror pairs.

We additionally define the median timing asymmetry

\begin{equation}
A_{t,\rm med}
=
\mathrm{median}
\left(
\Delta t
\right),
\end{equation}

which provides a complementary measure less sensitive to outliers.

The quantity $A_{t,\rm rms}$ measures the overall magnitude of the directional timing asymmetry, while $A_{t,\rm med}$ captures any net signed bias between opposite sides of the image plane.

In the Schwarzschild limit, reflection symmetry implies

\begin{equation}
\Delta t(\alpha,\beta)=0,
\end{equation}

such that both asymmetry observables vanish to numerical precision. In rotating Kerr spacetimes, the mirror-paired timing differences become systematically nonzero, leading to finite values of both asymmetry observables.

The observable introduced here therefore provides a geometric timing signature of Kerr rotation arising directly from relativistic photon propagation rather than detailed radiative transfer assumptions.

\subsection{Sampling and Robustness}

In constructing the asymmetry observable, we include only mirror pairs for which both trajectories successfully intersect the emitting surface. This prevents incomplete image-plane sampling from biasing the asymmetry measurement.

We additionally track the number of valid mirror pairs and the image-plane coverage fraction in each calculation to verify the statistical robustness of the measured signal.

Additional convergence and numerical-validation tests are presented in Section~\ref{sec:appendixB}.

% ---------- 6. Results ----------
\section{Results}\label{sec:results}

We now present the behavior of the time-delay asymmetry defined in Section~\ref{sec:asymmetry}, evaluated across a representative grid of black hole spin, observer inclination, and emission radius. The results presented below demonstrate how the mirror-paired timing asymmetry varies with Kerr rotation and viewing geometry and illustrate the emergence of systematic parity-breaking propagation effects in rotating spacetime.

\subsection{Schwarzschild Baseline}

We first consider the nonrotating Schwarzschild case ($a=0$) as a validation test of the symmetry properties of the numerical pipeline. Representative Schwarzschild time-delay maps are shown in Figure~\ref{fig:delay_maps}. As expected from spherical symmetry, the arrival-time structure remains reflection symmetric about the projected spin axis to numerical precision.

Consistent with this expectation, both asymmetry observables vanish:
\begin{equation}
A_{t,\rm rms} \approx 0,
\qquad
A_{t,\rm med} \approx 0.
\end{equation}

Residual deviations decrease with increasing numerical resolution and remain consistent with finite-resolution effects. This provides a stringent validation test confirming that the asymmetry observable isolates genuine spin-dependent propagation effects rather than numerical artifacts.

\subsection{Emergence of Asymmetry in Kerr Spacetime}

For rotating black holes ($a\neq0$), the time-delay maps exhibit a clear breakdown of reflection symmetry across the observer image plane. Representative Kerr delay maps are shown in Figure~\ref{fig:delay_maps}, where the asymmetry becomes progressively more pronounced with increasing spin.

The mirror-paired timing differences,

\begin{equation}
\Delta t(\alpha,\beta),
\end{equation}

become systematically nonzero across the image plane. Physically, this effect arises because frame dragging modifies the propagation times of prograde and retrograde photon trajectories differently \citep{Bardeen1972}.

The resulting asymmetry is visible both locally in the delay maps and globally through the scalar observables $A_{t,\rm rms}$ and $A_{t,\rm med}$.

\subsection{Dependence on Black Hole Spin}

\begin{figure}[t]
    \centering
    \includegraphics[width=\columnwidth]{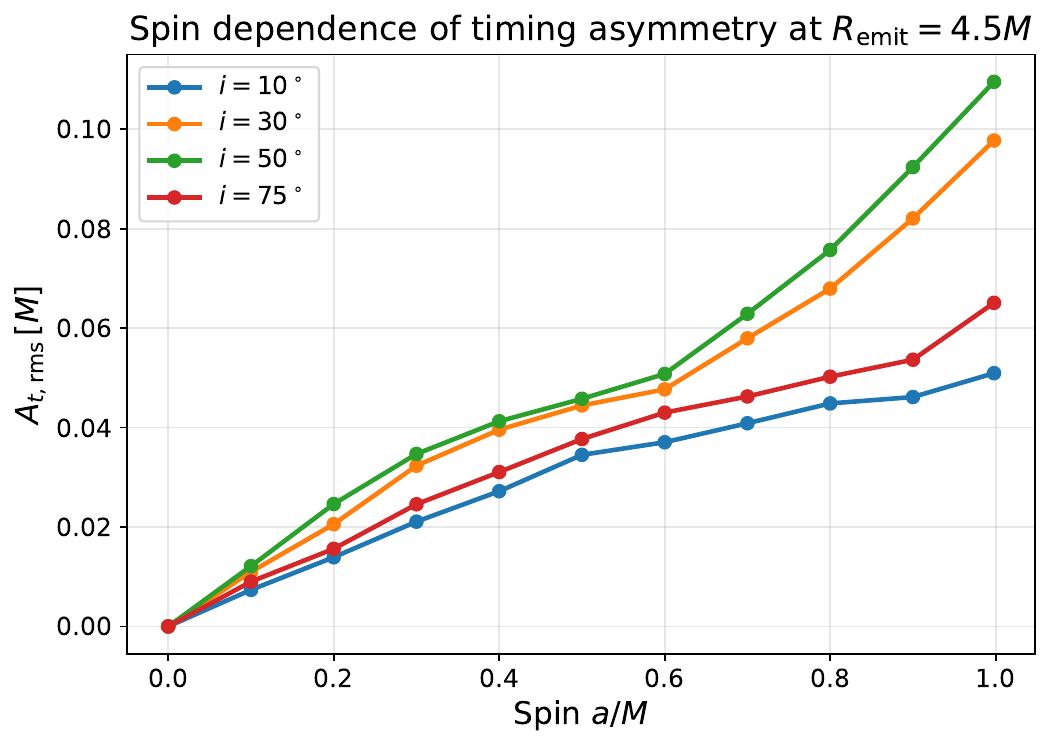}
    \caption{
    Root-mean-square timing asymmetry $A_{t,\rm rms}$ as a function of black hole spin for several observer inclinations at fixed emission radius $R_{\rm emit}=4.5M$.
    The asymmetry increases systematically with spin for all inclinations considered, demonstrating that the observable directly traces the influence of Kerr frame dragging.
    The strongest signals occur at intermediate-to-high inclinations, while nearly face-on configurations remain comparatively symmetric.
    The rapid increase toward high spin reflects the growing importance of strong-field relativistic photon propagation near the rotating black hole.
    }
    \label{fig:spin_dependence}
\end{figure}

Figure~\ref{fig:spin_dependence} shows the dependence of the root-mean-square timing asymmetry on black hole spin for several observer inclinations at fixed emission radius. The signal increases systematically with spin for all inclinations considered, demonstrating that the observable directly traces the growing influence of Kerr frame dragging on photon propagation.

For slowly rotating black holes, the timing asymmetry remains small but nonzero. As the spin approaches the near-extremal regime, however, the signal grows rapidly and becomes substantially more pronounced. This behavior reflects the increasing distortion of photon trajectories and propagation times in the strong-field region surrounding rapidly rotating black holes.

The inclination dependence further demonstrates that the observable is strongly geometry dependent. Nearly face-on systems remain comparatively symmetric, while intermediate-to-high inclination configurations exhibit significantly larger amplitudes. This trend indicates that the timing signature is sensitive not only to black hole spin itself, but also to the projected geometry of relativistic photon propagation across the observer image plane.

Overall, the systematic growth of $A_{t,\rm rms}$ with increasing spin establishes the timing asymmetry as a robust quantitative signature of Kerr spacetime and relativistic frame dragging.

\subsection{Inclination Dependence}

\begin{figure}[t]
    \centering
    \includegraphics[width=\columnwidth]{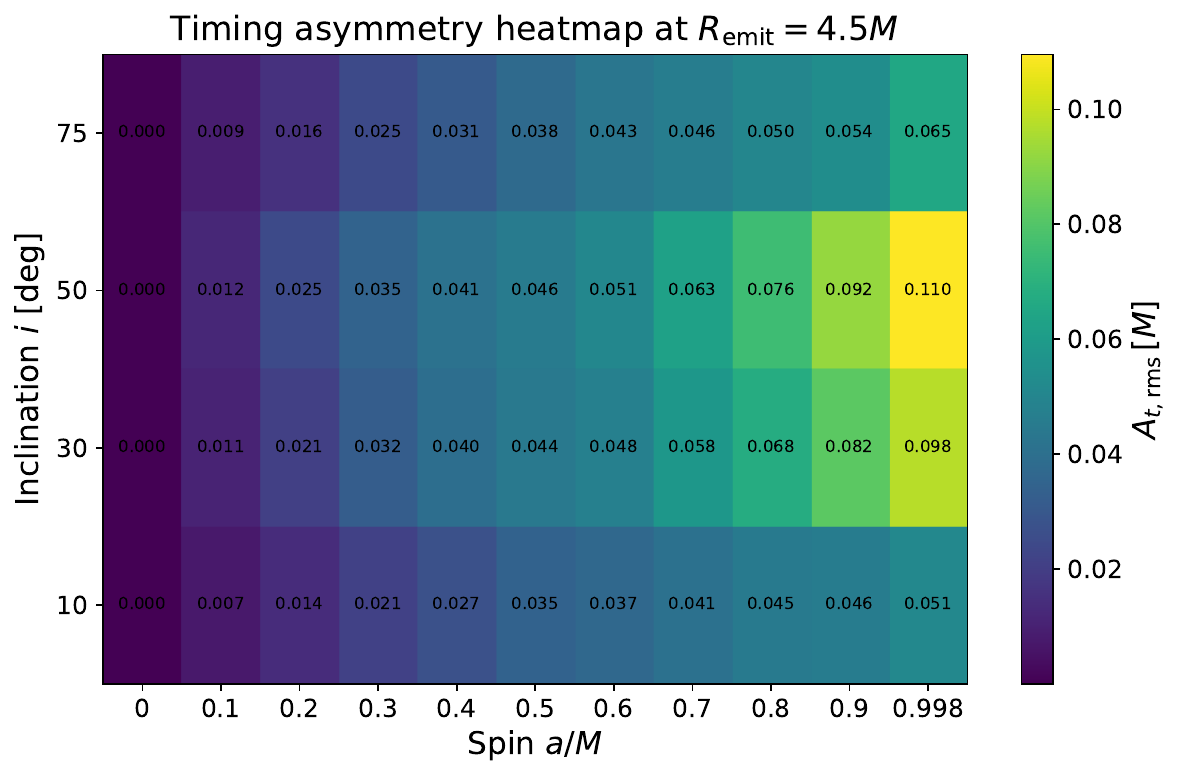}
    \caption{
    Heatmap of the root-mean-square timing asymmetry $A_{t,\rm rms}$ as a function of black hole spin and observer inclination for fixed emission radius $R_{\rm emit}=4.5M$.
    The asymmetry increases systematically with spin and exhibits a strong dependence on viewing geometry.
    Nearly face-on configurations remain comparatively symmetric, while intermediate-to-high inclinations produce substantially larger signal amplitudes.
    The strongest signals occur for rapidly rotating black holes viewed away from the face-on limit, reflecting the combined influence of frame dragging and relativistic photon propagation geometry.
    }
    \label{fig:heatmap}
\end{figure}

Figure~\ref{fig:heatmap} shows the dependence of the root-mean-square timing asymmetry on both black hole spin and observer inclination at fixed emission radius. The asymmetry increases systematically across the parameter space, with the strongest signals occurring for rapidly rotating black holes viewed at intermediate-to-high inclinations.

For nearly face-on systems, the projected image-plane geometry remains approximately symmetric, suppressing the observable and producing comparatively small values of $A_{t,\rm rms}$. As the inclination increases, however, the signal becomes significantly more pronounced due to the growing directional imbalance between photon trajectories propagating on opposite sides of the rotating spacetime.

Interestingly, the strongest asymmetries do not occur strictly in the edge-on limit, but instead emerge across intermediate-to-high inclinations. This behavior suggests that the observable depends not only on simple projection effects, but also on the detailed interplay between frame dragging, lensing geometry, and relativistic photon propagation near the black hole \citep{BeckwithDone2005}.

The heatmap structure further demonstrates that the timing asymmetry is jointly sensitive to both spin and viewing geometry, reinforcing its interpretation as a geometric signature of Kerr spacetime.

\subsection{Radial Dependence}

\begin{figure}[t]
    \centering
    \includegraphics[width=\columnwidth]{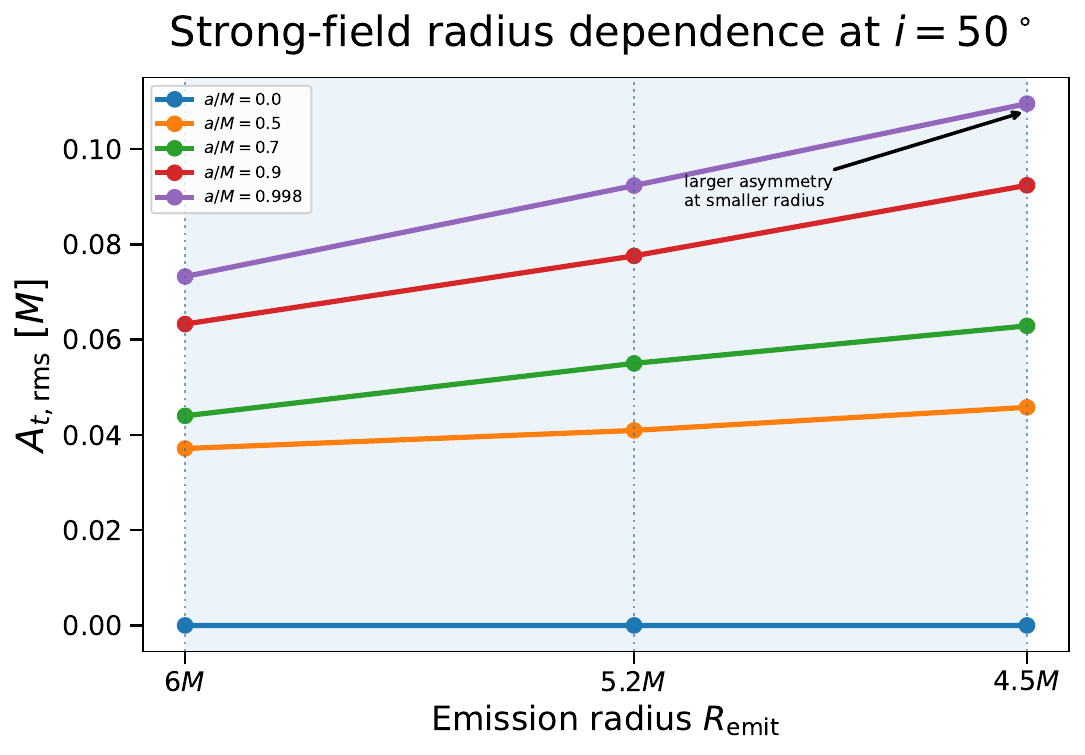}
    \caption{
    Root-mean-square timing asymmetry $A_{t,\rm rms}$ as a function of emission radius for several black hole spins at fixed observer inclination $i=50^\circ$.
    The vertical dotted lines mark the sampled emission radii $R_{\rm emit}=6M$, $5.2M$, and $4.5M$.
    The asymmetry increases as the emission radius moves inward, demonstrating that the signal is enhanced deeper in the relativistic strong-field regime.
    The trend is strongest for rapidly rotating black holes, while the Schwarzschild case remains consistent with zero.
    }
    \label{fig:radius_dependence}
\end{figure}

Figure~\ref{fig:radius_dependence} shows the dependence of the timing asymmetry on emission radius for several black hole spins at fixed inclination. Across the sampled radii, the asymmetry increases as $R_{\rm emit}$ decreases from $6M$ to $4.5M$, indicating that the signal becomes stronger closer to the black hole.

This radial trend supports the interpretation that the observable originates from relativistic photon propagation near the black hole. Close to the hole, frame dragging produces larger differences between mirror-paired photon trajectories, while at larger radii the effect weakens as the spacetime curvature decreases.

The increase is most pronounced for high-spin configurations, where frame dragging is strongest. In contrast, the Schwarzschild case remains consistent with zero across the explored radii, confirming that the radial dependence is associated with Kerr spin-induced symmetry breaking rather than numerical sampling effects.

\subsection{Representative Timing-Asymmetry Measurements}

Representative sequences are evaluated using fiducial configurations chosen to isolate the dependence of the timing asymmetry on a single parameter at a time. In particular, the spin sequence adopts an intermediate observer inclination of $i=50^\circ$, where the signal is both clearly measurable and sufficiently removed from the nearly face-on and edge-on projection limits.

\begin{table*}
\centering
\caption{
Representative timing-asymmetry measurements from the final $N=50$ production runs.
The table lists selected spin, inclination, and emission-radius sequences used to summarize
the dependence of the mirror-paired timing asymmetry on Kerr spin and viewing geometry.
}
\label{tab:main_results_summary}
\begin{tabular}{lcccccc}
\hline
Sequence & $a/M$ & $i$ [deg] & $R_{\rm emit}/M$
& $A_{t,\rm med}$ [$M$]
& $A_{t,\rm rms}$ [$M$]
& Hit fraction \\
\hline
Spin sequence & 0     & 50 & 4.5 & $-4.381\times10^{-10}$ & $3.955\times10^{-7}$ & 0.939 \\
Spin sequence & 0.3   & 50 & 4.5 & $-4.954\times10^{-3}$  & $3.470\times10^{-2}$ & 0.937 \\
Spin sequence & 0.5   & 50 & 4.5 & $-2.077\times10^{-3}$  & $4.579\times10^{-2}$ & 0.938 \\
Spin sequence & 0.7   & 50 & 4.5 & $4.615\times10^{-2}$   & $6.289\times10^{-2}$ & 0.935 \\
Spin sequence & 0.9   & 50 & 4.5 & $7.447\times10^{-2}$   & $9.241\times10^{-2}$ & 0.938 \\
Spin sequence & 0.998 & 50 & 4.5 & $9.303\times10^{-2}$   & $1.096\times10^{-1}$ & 0.936 \\
Inclination sequence & 0.998 & 10 & 4.5 & $4.747\times10^{-2}$ & $5.094\times10^{-2}$ & 0.942 \\
Inclination sequence & 0.998 & 30 & 4.5 & $8.149\times10^{-2}$ & $9.775\times10^{-2}$ & 0.940 \\
Inclination sequence & 0.998 & 75 & 4.5 & $4.760\times10^{-2}$ & $6.509\times10^{-2}$ & 0.933 \\
Radius sequence & 0.998 & 50 & 5.2 & $7.826\times10^{-2}$ & $9.234\times10^{-2}$ & 0.982 \\
Radius sequence & 0.998 & 50 & 6.0 & $5.521\times10^{-2}$ & $7.325\times10^{-2}$ & 1.000 \\
\hline
\end{tabular}
\end{table*}

Table~\ref{tab:main_results_summary} summarizes representative values of the timing-asymmetry observables from the final production runs. The spin sequence demonstrates the systematic growth of $A_{t,\rm rms}$ from the Schwarzschild baseline to the near-extremal Kerr regime. The inclination and radius sequences further show that the signal depends strongly on viewing geometry and emission location, with larger amplitudes generally occurring for smaller emission radii and intermediate inclinations.

Because $A_{t,\rm med}$ preserves the sign of the mirror-paired timing difference, its value depends on the adopted image-plane orientation and subtraction convention. In contrast, $A_{t,\rm rms}$ measures the overall amplitude of the asymmetry independent of sign.
\begin{comment}
Table~\ref{tab:main_results_summary} summarizes representative values of the timing-asymmetry observables from the final production runs. The spin sequence demonstrates the systematic growth of $A_{t,\rm rms}$ from the Schwarzschild baseline to the near-extremal Kerr case. The inclination and radius sequences show that the asymmetry is also strongly affected by viewing geometry and emission location, with larger values generally occurring for inner emission radii and intermediate inclinations.
\end{comment}

\subsection{Prograde and Retrograde Timing Structure}

\begin{figure}[t]
    \centering
    \includegraphics[width=\columnwidth]{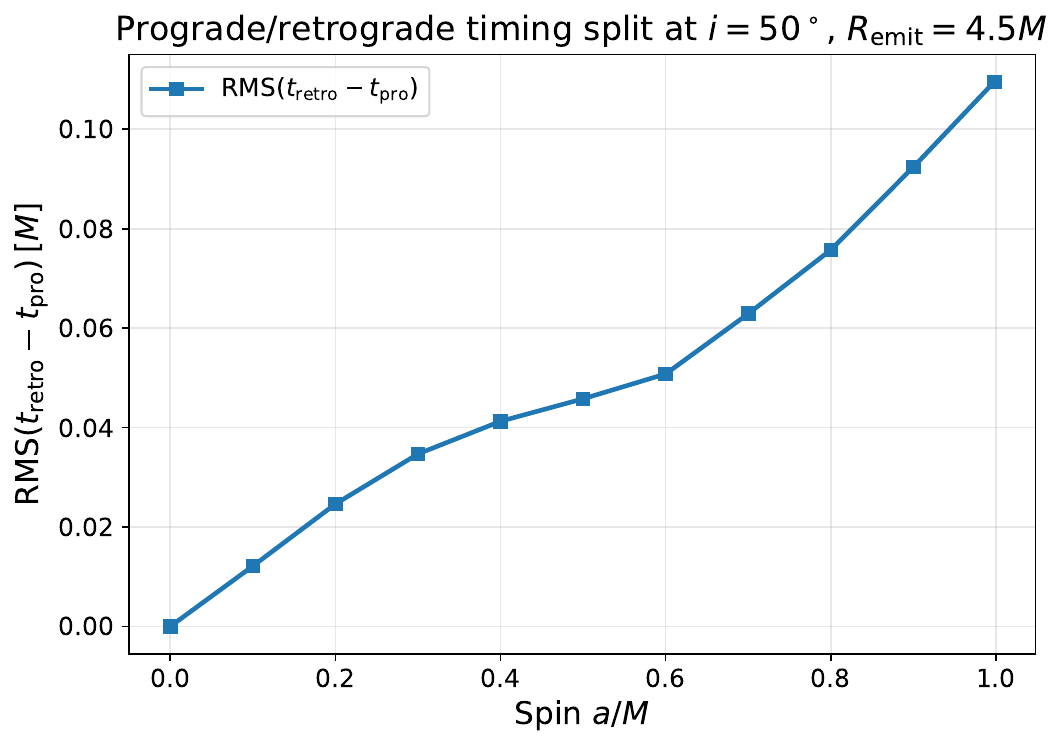}
    \caption{
    Prograde and retrograde contributions to the timing asymmetry as a function of black hole spin.
    The separation between the two trajectory classes increases with spin, demonstrating that the mirror-paired asymmetry is tied to directional photon propagation in Kerr spacetime.
    This behavior provides direct evidence that frame dragging modifies the travel times of photons on opposite sides of the projected spin axis.
    }
    \label{fig:prograde_retrograde}
\end{figure}

Figure~\ref{fig:prograde_retrograde} separates the timing structure into prograde and retrograde contributions. The difference between the two trajectory classes increases systematically with spin, confirming that the measured asymmetry is not simply a scalar numerical offset, but instead reflects directional photon propagation in the rotating spacetime.

This decomposition provides a direct physical interpretation of the observable. In Kerr spacetime, frame dragging modifies photon trajectories differently depending on whether they propagate with or against the sense of black hole rotation. As a result, prograde and retrograde paths accumulate different coordinate travel times before reaching the observer.

The prograde--retrograde separation therefore supports the interpretation that the mirror-paired timing asymmetry is a genuine frame-dragging signature rather than a numerical artifact or a generic property of curved spacetime.

\subsection{Astrophysical Scaling}
\label{subsec:astrophysical_scaling}

Although the timing asymmetry is computed in geometrized units, the corresponding physical timescales scale linearly with black hole mass through the gravitational timescale

\begin{equation}
1M = \frac{GM}{c^3}.
\end{equation}

The physical timing offset may therefore be obtained by multiplying the dimensionless asymmetry by the gravitational timescale associated with the black hole mass. This produces a simple linear scaling relation between the observable timing asymmetry and $M_{\rm BH}$.

The approximate scaling behavior of the asymmetry may also be understood from a simple order-of-magnitude argument. 
The mirror-paired timing asymmetry arises from spin-dependent modifications to null geodesic propagation near the black hole, producing fractional differences in photon travel times between trajectories propagating on opposite sides of the projected spin axis. 
Because the natural relativistic timescale of the spacetime is set by

\begin{equation}
t_g \sim \frac{GM}{c^3},
\end{equation}

the corresponding physical timing asymmetry scales approximately as

\begin{equation}
\Delta t_{\rm phys}
\sim
A_t
\frac{GM}{c^3},
\end{equation}

where $A_t$ represents the dimensionless geometrized asymmetry measured in units of $M$. 
As a result, even modest dimensionless asymmetries can correspond to astrophysically significant timing offsets for sufficiently massive black holes.

\begin{figure*}
\centering
\includegraphics[width=0.95\textwidth]{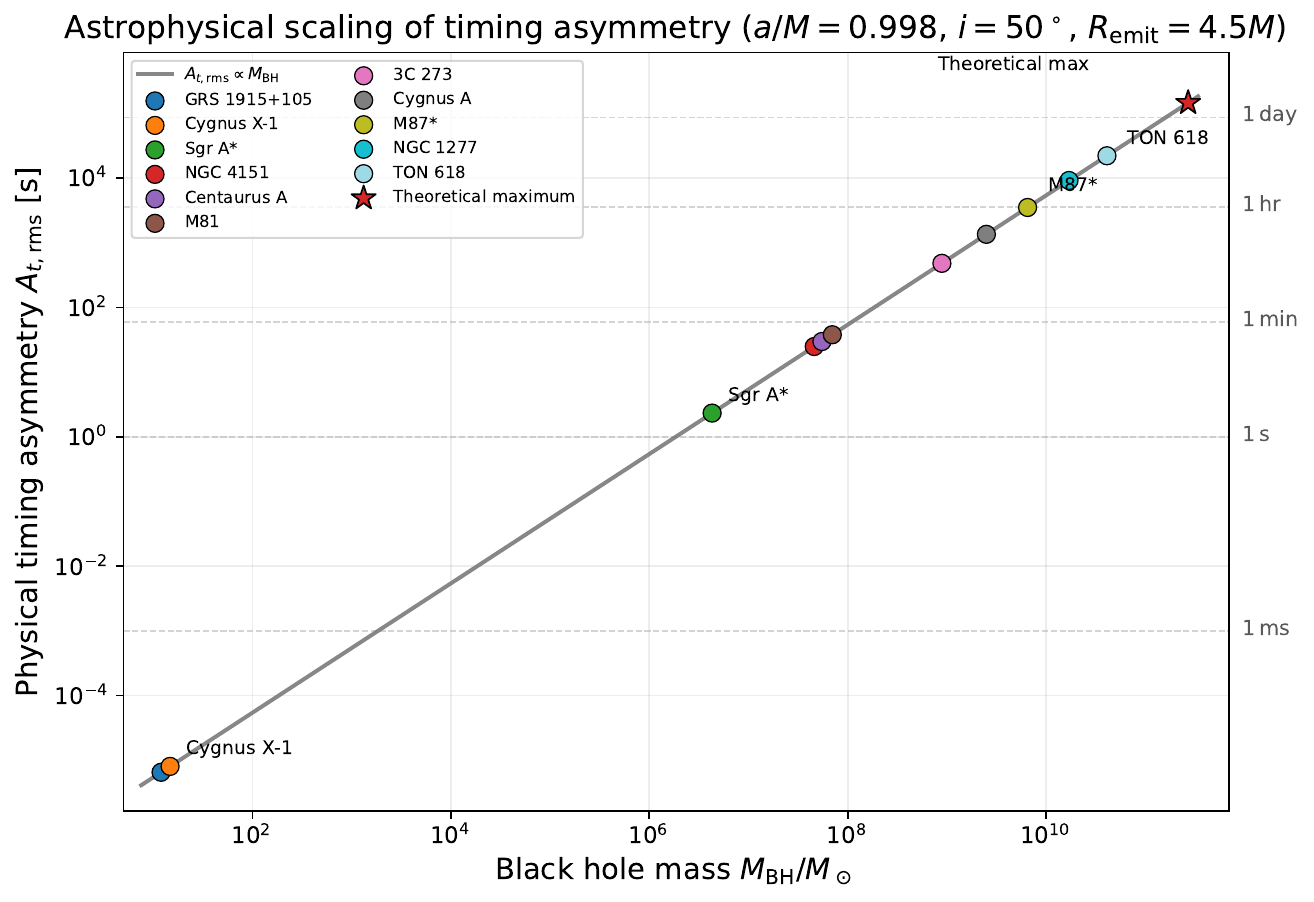}
\caption{
Physical scaling of the mirror-paired timing asymmetry for representative astrophysical black holes spanning stellar-mass X-ray binaries, supermassive black holes, and ultramassive quasars. The plotted values correspond to a representative high-spin Kerr configuration with $a/M=0.998$, $i=50^\circ$, and $R_{\rm emit}=4.5M$. The selected systems are shown to illustrate the mass scaling of the asymmetry rather than object-specific spin measurements. The asymmetry scales linearly with black hole mass through the gravitational timescale $GM/c^3$, producing microsecond-scale timing asymmetries for stellar-mass systems and hour- to day-scale asymmetries for ultramassive black holes.
}
\label{fig:astrophysical_scaling}
\end{figure*}

Figure~\ref{fig:astrophysical_scaling} shows the resulting physical scaling of the mirror-paired timing asymmetry for representative astrophysical black holes spanning stellar-mass X-ray binaries, nearby supermassive black holes, and ultramassive quasars. The displayed values correspond to a representative high-spin Kerr configuration with $a/M=0.998$, $i=50^\circ$, and $R_{\rm emit}=4.5M$. The selected systems are shown to illustrate the mass scaling of the asymmetry rather than object-specific spin measurements.

\begin{table*}
\centering
\caption{
Representative astrophysical black holes adopted for the timing-asymmetry scaling analysis.
The listed masses are taken from the literature and are intended to illustrate the physical scaling of the asymmetry across stellar-mass and supermassive black hole systems.
}
\label{tab:bh_masses}

\begin{tabular}{lcc}
\hline
Object & $M_{\rm BH}\,[M_\odot]$ & Reference \\
\hline
GRS 1915+105 & $1.2\times10^{1}$ & 1 \\
Cygnus X-1 & $1.5\times10^{1}$ & 2 \\
Sgr A* & $4.3\times10^{6}$ & 3 \\
NGC 4151 & $4.6\times10^{7}$ & 4 \\
Centaurus A & $5.5\times10^{7}$ & 5 \\
M81 & $7.0\times10^{7}$ & 6 \\
3C 273 & $8.9\times10^{8}$ & 7 \\
Cygnus A & $2.5\times10^{9}$ & 8 \\
M87$^\ast$ & $6.5\times10^{9}$ & 9 \\
NGC 1277 & $1.7\times10^{10}$ & 10 \\
TON 618 & $4.1\times10^{10}$ & 11 \\
\hline
\end{tabular}

\vspace{0.2cm}

\textbf{References.}
(1) \citep{Reid2014};
(2) \citep{Orosz2011};
(3) \citep{Gravity2019};
(4) \citep{Bentz2006};
(5) \citep{Neumayer2010};
(6) \citep{Devereux2003};
(7) \citep{Paltani2005};
(8) \citep{Tadhunter2003};
(9) \citep{EHT2019_VI};
(10) \citep{vanDenBosch2012};
(11) \citep{Ge2019}.
\end{table*}

Table~\ref{tab:bh_masses} summarizes representative black hole masses adopted in the scaling analysis.

For stellar-mass black holes such as Cygnus~X-1 and GRS~1915+105, the predicted asymmetry corresponds to microsecond- to millisecond-scale timing differences. In contrast, supermassive black holes produce second-, minute-, and hour-scale asymmetries, while the most massive known quasars approach day-scale timing offsets.

The approximately linear trend observed in log--log space reflects the direct proportionality between the gravitational timescale and black hole mass. While the precise numerical values depend on the adopted spin, inclination, and emission geometry, the overall scaling behavior is a generic consequence of relativistic photon propagation in Kerr spacetime.

These results suggest that spin-induced directional timing asymmetries may become observationally relevant for future high-precision reverberation mapping and strong-gravity timing studies of accreting black holes \citep{Reynolds2021}. In particular, rapidly rotating supermassive black holes provide the most favorable regime for detecting measurable directional timing signatures associated with Kerr frame dragging.

\subsection{Numerical Robustness}

We verify that the measured timing asymmetry remains stable with respect to both image-plane resolution and geodesic integration step size. Convergence tests demonstrate that the asymmetry observables $A_{t,\rm rms}$ and $A_{t,\rm med}$ approach stable values as the numerical resolution is increased, indicating that the measured signal is not dominated by discretization effects.

We additionally track the number and fraction of valid mirror pairs contributing to each calculation. The resulting hit fractions remain consistently high across the explored parameter space, demonstrating that the measured asymmetry is not produced by sparse sampling or localized numerical artifacts.

The Schwarzschild limit further provides an important validation test: for $a=0$, the asymmetry vanishes to numerical precision, confirming that the observed signal originates from Kerr spin-induced symmetry breaking rather than numerical bias.

\subsection{Summary of Results}

The principal results of this work can be summarized as follows:

\begin{enumerate}
    \item The timing asymmetry vanishes in the Schwarzschild limit, providing a strong validation test of both the observable and the numerical framework.
    
    \item Rotating Kerr spacetimes produce a nonzero mirror-paired timing asymmetry through relativistic frame dragging.
    
    \item The asymmetry increases systematically with black hole spin, becoming most pronounced in the near-extremal regime.
    
    \item The effect depends strongly on viewing geometry and emission location, with the largest asymmetries occurring for intermediate-to-high inclinations and emission originating deeper in the strong-field region.
    
    \item The corresponding physical timing offsets scale linearly with black hole mass and reach astrophysically significant magnitudes for supermassive black holes.
\end{enumerate}

These results establish the mirror-paired timing asymmetry as a robust and physically interpretable observable encoding information about relativistic photon propagation and Kerr spacetime geometry.

% ---------- 7. Discussion and Observational Implications ----------
\section{Discussion and Observational Implications}
\label{sec:discussion}

The results presented in this work demonstrate that Kerr frame dragging produces a systematic breaking of reflection symmetry in photon arrival times across the observer image plane. By constructing mirror-paired timing differences directly from relativistic time-delay maps, we isolate a geometric timing signature that vanishes in the Schwarzschild limit and grows systematically with black hole spin.

A key feature of the observable introduced here is that it preserves directional information across the image plane. Many traditional timing analyses rely on spatially integrated transfer functions or averaged temporal signals \citep{WilkinsFabian2012,Uttley2014,Emmanoulopoulos2014}, which can partially wash out asymmetries associated with relativistic photon propagation. In contrast, the mirror-paired construction isolates differences between photons arriving from opposite sides of the projected spin axis, allowing the underlying frame-dragging geometry to be probed more directly.

The observed asymmetry originates fundamentally from the differential propagation of prograde and retrograde photon trajectories in Kerr spacetime \citep{Bardeen1972,BeckwithDone2005}. Frame dragging modifies both the effective path length and coordinate propagation time of photons traveling near the rotating black hole, producing systematic differences in arrival-time structure across the image plane. The resulting asymmetry therefore represents a direct manifestation of relativistic photon propagation in rotating spacetime.

Our results also reveal a nontrivial dependence on observer inclination. While the asymmetry generally increases as the system departs from a face-on geometry, the strongest signals occur at intermediate-to-high inclinations rather than in strictly edge-on configurations. This behavior suggests that the observable depends not only on simple projection effects, but also on the detailed interplay between frame dragging, lensing geometry, and photon trajectory structure near the black hole.

The radial dependence of the asymmetry further supports its strong-field origin. Emission originating close to the black hole produces substantially larger asymmetries than emission at larger radii, consistent with the increasing influence of frame dragging in the inner strong-field region. This behavior indicates that the observable is most sensitive to photon trajectories probing the deepest parts of the Kerr spacetime.

An important aspect of this work is that the observable is defined geometrically and does not rely on detailed assumptions regarding radiative transfer, emissivity profiles, or plasma microphysics. While realistic astrophysical systems will inevitably introduce additional complexities, the underlying timing asymmetry arises directly from the spacetime geometry itself. This makes the observable conceptually robust and potentially useful as a complementary probe of black hole spin \citep{Reynolds2021}.

At the same time, several limitations should be emphasized. The present study adopts an idealized thin-disk geometry and does not include effects such as disk thickness, turbulence, magnetic fields, scattering, or time-dependent plasma variability. In realistic accretion environments, these processes may alter or partially obscure the asymmetry signal. Future work incorporating more realistic emission geometries and radiative-transfer effects will therefore be important for assessing the full observational implications of the phenomenon.

Despite these limitations, the astrophysical scaling presented in Section~\ref{sec:results} suggests that the timing asymmetry can reach physically significant scales for supermassive black holes. For sufficiently massive systems, the predicted timing offsets extend from seconds to hours, indicating that directional timing asymmetries may become relevant for future high-precision reverberation and strong-gravity timing studies \citep{Uttley2014,Reynolds2021}.

More broadly, the results presented here suggest that spatially resolved timing structure may encode additional information about relativistic photon propagation beyond that captured by conventional integrated timing observables. Future high-resolution timing and imaging studies may therefore provide new opportunities to probe directional signatures of strong-field gravity in the vicinity of rotating black holes.

\subsection{Connection to Reverberation Mapping}

Time delays between photons emitted from different regions of an accretion flow are central observables in X-ray reverberation mapping \citep{Reynolds1999,Uttley2014}. In such studies, the temporal response of reflected emission encodes information about the geometry and relativistic structure of the inner accretion disk.

The mirror-paired timing asymmetry introduced here adds a directional component to this timing structure. Photons emitted from opposite sides of the disk relative to the projected spin axis experience systematically different propagation times, producing an anisotropic contribution to the arrival-time distribution.

In principle, such asymmetries may manifest as directional skewness or asymmetric structure within relativistic transfer functions \citep{WilkinsFabian2012,Emmanoulopoulos2014}, particularly in systems where the observed emission is dominated by compact inner-disk regions.

\subsection{Dependence on System Geometry}

The timing asymmetry is strongest for systems viewed at intermediate-to-high inclinations and for emission originating close to the black hole. These conditions maximize the influence of frame dragging and relativistic photon propagation \citep{Bardeen1972,BeckwithDone2005}.

Compact corona--disk geometries and systems exhibiting strong inner-disk reflection are therefore likely to provide the most favorable environments for the appearance of the effect.

Figure~\ref{fig:astrophysical_scaling} further illustrates that many observationally studied high-spin black hole systems occupy the parameter regime where the predicted asymmetry becomes most pronounced. Although the present calculations adopt an idealized fiducial geometry and do not attempt source-specific modeling, the comparison demonstrates that the observable naturally arises within physically relevant regions of Kerr parameter space.

\subsection{Observational Challenges}

Detecting directional timing asymmetries observationally remains challenging. Current X-ray timing observations typically probe spatially integrated signals across the accretion flow, which can dilute the asymmetry through averaging over azimuthal structure and emission geometry.

Additional uncertainties associated with coronal structure, emissivity profiles, turbulence, and time-dependent variability may further complicate extraction of the signal. Furthermore, the predicted asymmetry must ultimately be assessed relative to the timing resolution and signal-to-noise ratios achievable with current and future instruments.

For these reasons, the present work should primarily be viewed as establishing the existence and physical behavior of the asymmetry rather than claiming immediate detectability in existing observational datasets.

\subsection{Future Prospects}

Future timing observatories with improved sensitivity and higher temporal resolution may enable more detailed reconstruction of relativistic arrival-time distributions. In particular, advances in reverberation analysis, timing tomography, and strong-gravity imaging techniques may eventually make it possible to probe directional asymmetries associated with Kerr frame dragging \citep{Uttley2014,Reynolds2021}.

The timing asymmetry explored here may also complement other relativistic observables, including spectral reflection features, horizon-scale imaging constraints, and polarization measurements \citep{Miller2007,SchnittmanKrolik2010,Reynolds2021}. Combined analyses using multiple observables could help break degeneracies between black hole spin, inclination, and emission geometry.

If measurable, the mirror-paired timing asymmetry would provide a new geometric probe of black hole rotation arising directly from relativistic photon propagation in Kerr spacetime.

% ---------- 9. Conclusions ----------
\section{Conclusions}\label{sec:conclusion}

We have introduced a new timing observable designed to probe spin-induced asymmetries in relativistic photon propagation near rotating black holes. Using backward ray tracing in Kerr spacetime \citep{Kerr1963}, we constructed image-plane time-delay maps and quantified the resulting mirror-paired timing asymmetry between photons arriving from opposite sides of the projected spin axis.

The asymmetry vanishes in the Schwarzschild limit and becomes systematically nonzero in rotating Kerr spacetimes, confirming its origin in frame dragging and spin-dependent photon propagation \citep{Bardeen1972}. The signal increases with black hole spin and depends strongly on both observer inclination and emission radius, with the largest effects arising in high-spin configurations and within the inner strong-field region.

Scaling the dimensionless asymmetry to physical units shows that the corresponding timing offsets range from microseconds for stellar-mass systems to hour-scale delays for supermassive black holes. These results suggest that directional timing structure may become relevant for future strong-gravity timing and reverberation studies \citep{Reynolds1999,Uttley2014,Cackett2021}.

More broadly, this work demonstrates that symmetry-based analyses of image-plane timing structure can reveal new information about relativistic photon propagation in Kerr spacetime. The mirror-paired timing asymmetry introduced here provides a geometrically motivated complement to existing spectral, imaging, polarization, and timing probes of black hole spin \citep{Cunningham1975,Laor1991,Miller2007,SchnittmanKrolik2010,Reynolds2021}.

Future work incorporating realistic radiative transfer, extended emission geometries, and time-dependent accretion physics will be important for assessing the observational implications of the effect. Nevertheless, the present results establish the timing asymmetry as a robust theoretical signature of relativistic photon propagation in rotating spacetime.
% ---------- Acknowledgments ----------
\section*{Acknowledgments}

The author thanks his family for their continued support and encouragement.
The author is especially grateful to his wife for her support throughout this work.
The author also sincerely thanks Prof. Joshua Tan for his encouragement and support of this independent work.

% ---------- References ----------
\bibliographystyle{aasjournal}
\bibliography{refs}

@article{SchnittmanKrolik2010,
  author = {Schnittman, Jeremy D. and Krolik, Julian H.},
  title = {X-Ray Polarization from Accreting Black Holes: Coronal Emission},
  journal = {The Astrophysical Journal},
  volume = {712},
  number = {2},
  pages = {908--924},
  year = {2010},
  doi = {10.1088/0004-637X/712/2/908}
}

@article{Cunningham1975,
  author = {Cunningham, Charles T.},
  title = {The Effects of Redshifts and Focusing on the Spectrum of an Accretion Disk around a Kerr Black Hole},
  journal = {The Astrophysical Journal},
  volume = {202},
  pages = {788--802},
  year = {1975},
  doi = {10.1086/154033}
}

@article{Laor1991,
  author = {Laor, Ari},
  title = {Line Profiles from a Disk around a Rotating Black Hole},
  journal = {The Astrophysical Journal},
  volume = {376},
  pages = {90--94},
  year = {1991},
  doi = {10.1086/170257}
}

@article{Reynolds1999,
  author = {Reynolds, Christopher S. and Young, Andrew J. and Begelman, Mitchell C. and Fabian, Andrew C.},
  title = {X-ray Iron Line Reverberation from Black Hole Accretion Disks},
  journal = {The Astrophysical Journal},
  volume = {514},
  number = {1},
  pages = {164--179},
  year = {1999},
  doi = {10.1086/306932}
}

@article{Uttley2014,
  author = {Uttley, P. and Cackett, E. M. and Fabian, A. C. and Kara, E. and Wilkins, D. R.},
  title = {X-ray Reverberation around Accreting Black Holes},
  journal = {The Astronomy and Astrophysics Review},
  volume = {22},
  number = {1},
  pages = {72},
  year = {2014},
  doi = {10.1007/s00159-014-0072-0}
}

@article{Reynolds2021,
  author = {Reynolds, Christopher S.},
  title = {Observing Black Holes Spin},
  journal = {Annual Review of Astronomy and Astrophysics},
  volume = {59},
  pages = {117--154},
  year = {2021},
  doi = {10.1146/annurev-astro-112420-035022}
}

@article{WilkinsFabian2013,
  author = {Wilkins, D. R. and Fabian, A. C.},
  title = {Understanding X-ray Reflection Emissivity Profiles in AGN: Locating the X-ray Source},
  journal = {Monthly Notices of the Royal Astronomical Society},
  volume = {430},
  number = {4},
  pages = {247--258},
  year = {2013},
  doi = {10.1093/mnras/sts612}
}

@article{WilkinsFabian2012,
  author = {Wilkins, D. R. and Fabian, A. C.},
  title = {Understanding X-ray Reverberation around Accreting Black Holes},
  journal = {Monthly Notices of the Royal Astronomical Society},
  volume = {424},
  number = {2},
  pages = {1284--1296},
  year = {2012},
  doi = {10.1111/j.1365-2966.2012.21308.x}
}

@article{Emmanoulopoulos2014,
  author = {Emmanoulopoulos, D. and McHardy, I. M. and Papadakis, I. E.},
  title = {General Relativistic Impulse Response Functions of X-ray Reflection},
  journal = {Monthly Notices of the Royal Astronomical Society},
  volume = {404},
  number = {2},
  pages = {931--942},
  year = {2014},
  doi = {10.1111/j.1365-2966.2010.16342.x}
}

@article{BeckwithDone2005,
  author = {Beckwith, Kris and Done, Chris},
  title = {Extreme Gravitational Lensing near Rotating Black Holes},
  journal = {Monthly Notices of the Royal Astronomical Society},
  volume = {359},
  number = {4},
  pages = {1217--1228},
  year = {2005},
  doi = {10.1111/j.1365-2966.2005.08941.x}
}

@article{Bardeen1972,
  author = {Bardeen, James M. and Press, William H. and Teukolsky, Saul A.},
  title = {Rotating Black Holes: Locally Nonrotating Frames, Energy Extraction, and Scalar Synchrotron Radiation},
  journal = {The Astrophysical Journal},
  volume = {178},
  pages = {347--370},
  year = {1972},
  doi = {10.1086/151796}
}

@article{Miller2007,
  author = {Miller, Jon M.},
  title = {Relativistic X-Ray Lines from the Inner Accretion Disks around Black Holes},
  journal = {Annual Review of Astronomy and Astrophysics},
  volume = {45},
  pages = {441--479},
  year = {2007},
  doi = {10.1146/annurev.astro.45.051806.110555}
}

@article{Kerr1963,
  author = {Kerr, Roy P.},
  title = {Gravitational Field of a Spinning Mass as an Example of Algebraically Special Metrics},
  journal = {Physical Review Letters},
  volume = {11},
  number = {5},
  pages = {237--238},
  year = {1963},
  doi = {10.1103/PhysRevLett.11.237}
}

@article{Fanton1997,
  author = {Fanton, C. and Calvani, M. and de Felice, F. and Cadez, A.},
  title = {Detecting Accretion Disks in Active Galactic Nuclei},
  journal = {Publications of the Astronomical Society of Japan},
  volume = {49},
  number = {2},
  pages = {159--169},
  year = {1997},
  doi = {10.1093/pasj/49.2.159}
}

@article{Vincent2011,
  author = {Vincent, Fr{\'e}d{\'e}ric H. and Paumard, Thibault and Gourgoulhon, Eric and Perrin, Guy},
  title = {GYOTO: a New General Relativistic Ray-Tracing Code},
  journal = {Classical and Quantum Gravity},
  volume = {28},
  number = {22},
  pages = {225011},
  year = {2011},
  doi = {10.1088/0264-9381/28/22/225011}
}

@article{Orosz2011,
  author = {Orosz, Jerome A. and McClintock, Jeffrey E. and Aufdenberg, Jason P. and Remillard, Ronald A. and Reid, Mark J. and Narayan, Ramesh and Gou, Lijun},
  title = {The Mass of the Black Hole in Cygnus X-1},
  journal = {The Astrophysical Journal},
  volume = {742},
  number = {2},
  pages = {84},
  year = {2011},
  doi = {10.1088/0004-637X/742/2/84},
  eprint = {1106.3689},
  archivePrefix = {arXiv},
  primaryClass = {astro-ph.SR}
}

@article{EHT2019_V,
  author = {{Event Horizon Telescope Collaboration}},
  title = {First M87 Event Horizon Telescope Results. V. Physical Origin of the Asymmetric Ring},
  journal = {ApJL},
  volume = {875},
  pages = {L5},
  year = {2019}
}

@article{EHT2019_VI,
  author = {{Event Horizon Telescope Collaboration}},
  title = {First M87 Event Horizon Telescope Results. VI. The Shadow and Mass of the Central Black Hole},
  journal = {The Astrophysical Journal Letters},
  volume = {875},
  number = {1},
  pages = {L6},
  year = {2019},
  doi = {10.3847/2041-8213/ab1141},
  eprint = {1906.11243},
  archivePrefix = {arXiv},
  primaryClass = {astro-ph.GA}
}

@article{Gravity2019,
  author = {{GRAVITY Collaboration}},
  title = {A Geometric Distance Measurement to the Galactic Center Black Hole with 0.3\% Uncertainty},
  journal = {Astronomy \& Astrophysics},
  volume = {625},
  pages = {L10},
  year = {2019},
  doi = {10.1051/0004-6361/201935656},
  eprint = {1904.05721},
  archivePrefix = {arXiv},
  primaryClass = {astro-ph.GA}
}

@article{Reid2014,
  author = {Reid, Mark J. and others},
  title = {A Parallax Distance to the Microquasar GRS 1915+105 and a Revised Estimate of Its Black Hole Mass},
  journal = {The Astrophysical Journal},
  volume = {796},
  number = {1},
  pages = {2},
  year = {2014},
  doi = {10.1088/0004-637X/796/1/2},
  eprint = {1409.2453},
  archivePrefix = {arXiv},
  primaryClass = {astro-ph.GA}
}

@article{Bentz2006,
  author = {Bentz, Misty C. and others},
  title = {A Reverberation-Based Mass for the Central Black Hole in NGC 4151},
  journal = {The Astrophysical Journal},
  volume = {651},
  number = {2},
  pages = {775--781},
  year = {2006},
  doi = {10.1086/507589},
  eprint = {astro-ph/0607085},
  archivePrefix = {arXiv}
}

@article{vanDenBosch2012,
  author = {van den Bosch, Remco C. E. and others},
  title = {An Over-massive Black Hole in the Compact Lenticular Galaxy NGC 1277},
  journal = {Nature},
  volume = {491},
  pages = {729--731},
  year = {2012},
  doi = {10.1038/nature11592},
  eprint = {1211.6429},
  archivePrefix = {arXiv},
  primaryClass = {astro-ph.CO}
}

@article{Ge2019,
  author = {Ge, Xue and Zhao, Bi-Xuan and Bian, Wei-Hao and Green, Richard},
  title = {The Blueshift of the C IV Broad Emission Line in QSOs},
  journal = {The Astronomical Journal},
  volume = {157},
  number = {2},
  pages = {81},
  year = {2019},
  doi = {10.3847/1538-3881/ab01d2},
  eprint = {1903.11023},
  archivePrefix = {arXiv},
  primaryClass = {astro-ph.GA}
}

@article{Neumayer2010,
  author = {Neumayer, Nadine},
  title = {The Supermassive Black Hole at the Heart of Centaurus A: Revealed by Gas- and Star-Kinematic Measurements},
  journal = {Publications of the Astronomical Society of Australia},
  volume = {27},
  number = {4},
  pages = {449--456},
  year = {2010},
  doi = {10.1071/AS09057},
  eprint = {1001.5212},
  archivePrefix = {arXiv},
  primaryClass = {astro-ph.CO}
}

@article{Devereux2003,
  author = {Devereux, Nick and Ford, Holland and Tsvetanov, Zolt and Jacoby, George},
  title = {Hubble Space Telescope Evidence for an Intermediate-Mass Black Hole in M81},
  journal = {The Astronomical Journal},
  volume = {125},
  number = {3},
  pages = {1226--1234},
  year = {2003},
  doi = {10.1086/346139},
  eprint = {astro-ph/0211238},
  archivePrefix = {arXiv}
}

@article{Paltani2005,
  author = {Paltani, S. and Türler, M.},
  title = {The Black Hole Mass of 3C 273},
  journal = {Astronomy \& Astrophysics},
  volume = {435},
  number = {3},
  pages = {811--821},
  year = {2005},
  doi = {10.1051/0004-6361:20041998},
  eprint = {astro-ph/0503265},
  archivePrefix = {arXiv}
}

@article{Tadhunter2003,
  author = {Tadhunter, Clive and Marconi, Alessandro and Axon, David and Robinson, Andrew and Jackson, Neal and Villar-Martín, Montserrat},
  title = {The Black Hole Mass in Cygnus A},
  journal = {Monthly Notices of the Royal Astronomical Society},
  volume = {342},
  number = {3},
  pages = {861--878},
  year = {2003},
  doi = {10.1046/j.1365-8711.2003.06588.x},
  eprint = {astro-ph/0302231},
  archivePrefix = {arXiv}
}

@article{Gralla2019,
  author = {Gralla, Samuel E. and Holz, Daniel E. and Wald, Robert M.},
  title = {Black Hole Shadows, Photon Rings, and Lensing Rings},
  journal = {Phys. Rev. D},
  volume = {100},
  pages = {024018},
  year = {2019}
}

@article{Cackett2021,
  author = {Cackett, Edward M. et al.},
  title = {X-ray Reverberation around Accreting Black Holes},
  journal = {Space Science Reviews},
  volume = {217},
  pages = {57},
  year = {2021}
}

@article{Chowdhury2026,
  author = {Chowdhury, Shakibul},
  title = {Kerr Polarization Transport: Accuracy and Performance in General Relativistic Light Propagation},
  journal = {The Astrophysical Journal},
  year = {2026},
  doi = {10.3847/1538-4357/ae6652}
}

% ---------- Appendix ----------
\appendix

\section{Additional Numerical Tests}
\label{sec:appendixA}

To verify that the measured timing asymmetry arises from Kerr frame dragging rather than numerical artifacts, we performed several additional validation tests beyond the primary production calculations discussed in the main text.

First, Schwarzschild ($a=0$) delay maps were computed using the identical ray-tracing pipeline employed for the Kerr calculations. As expected from the reflection symmetry of Schwarzschild spacetime, the mirror-defined timing asymmetry remained consistent with zero within numerical precision. Typical values satisfied
\[
A_{t,\rm rms} \sim 10^{-8} - 10^{-10},
\]
depending on resolution and integration parameters.

We additionally verified that the sign-separated prograde and retrograde timing differences behaved consistently under changes in image-plane resolution and integration step size. The qualitative morphology of the delay maps remained stable across all tested configurations.

The orthogonality and null constraints associated with the photon four-momentum were monitored throughout the integrations. Numerical drift remained negligible over the full geodesic trajectories considered in this work.

\section{Convergence and Resolution Studies}
\label{sec:appendixB}

\begin{figure}[t]
    \centering
    \includegraphics[width=\columnwidth]{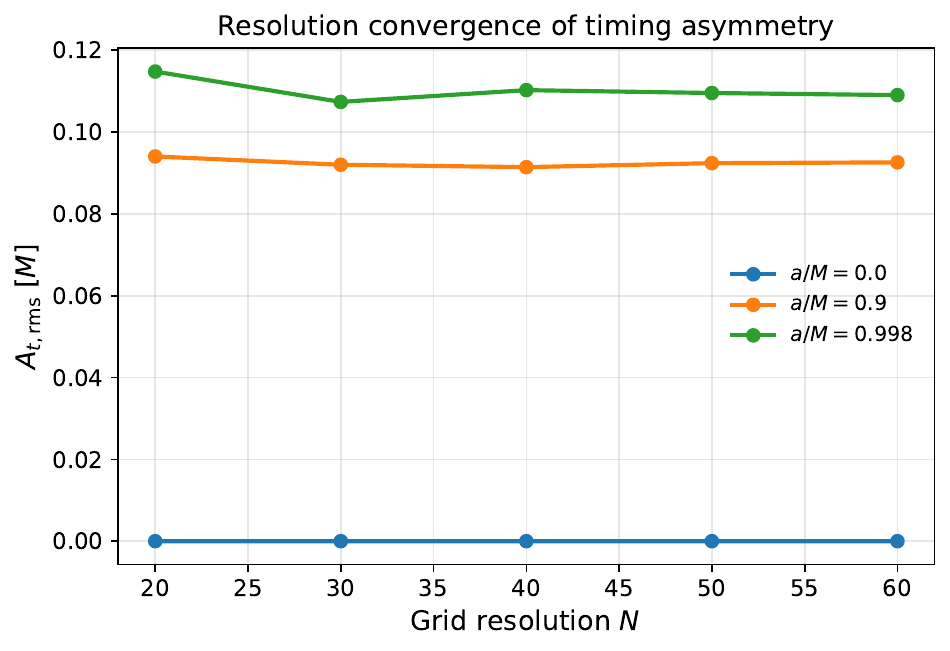}
    \caption{
    Resolution convergence of the root-mean-square timing asymmetry $A_{t,\rm rms}$ for representative Schwarzschild and Kerr configurations.
    The asymmetry stabilizes systematically with increasing image-plane resolution $N$, demonstrating that the measured signal is robust against finite sampling effects.
    The Schwarzschild case remains consistent with zero asymmetry across all tested resolutions, while the Kerr cases converge toward stable nonzero amplitudes.
    }
    \label{fig:convergence}
\end{figure}

To assess convergence, the timing-asymmetry calculations were repeated over a sequence of image-plane resolutions,
\[
N = 20,\ 30,\ 40,\ 50,\ 60,
\]
for representative low-spin and high-spin configurations.

The convergence tests were performed for selected combinations of spin, inclination, and emission radius, including both the Schwarzschild limit and near-extremal Kerr cases. Representative convergence curves are shown in Figure~\ref{fig:convergence}.

The resulting values of $A_{t,\rm rms}$ exhibited systematic stabilization with increasing resolution. In particular, the high-spin Kerr configurations converged toward stable nonzero asymmetry amplitudes as the image-plane sampling was refined. Differences between successive resolutions decreased monotonically, indicating that the measured asymmetry is not produced by undersampling of the observer screen or interpolation artifacts.

The Schwarzschild case remained consistent with zero asymmetry across all tested resolutions, further confirming that the observable isolates spin-induced directional propagation effects rather than numerical bias. Across the production calculations adopted in the main text, uncertainties associated with finite image resolution remained substantially smaller than the physical spin-dependent trends discussed in the Results section.

\section{Example Conversion to Physical Units}
\label{sec:appendix_scaling}

The timing asymmetry calculated throughout this work is expressed in geometrized units of $M$, where

\begin{equation}
1M = \frac{GM_{\rm BH}}{c^3}.
\end{equation}

To convert the asymmetry into physical units, the geometrized timing observable is multiplied by the gravitational timescale associated with the black hole mass.

For the representative high-spin configuration
\[
a/M = 0.998,
\qquad
i=50^\circ,
\qquad
R_{\rm emit}=4.5M,
\]
the measured root-mean-square asymmetry is approximately
\[
A_{t,\rm rms} \approx 0.1096M.
\]

Using
\[
\frac{GM_\odot}{c^3}
=
4.9255\times10^{-6}\ {\rm s},
\]
the physical timing asymmetry becomes
\begin{equation}
\Delta t_{\rm phys}
=
A_t
\frac{GM_{\rm BH}}{c^3}
\approx
5.40\times10^{-7}
\left(
\frac{M_{\rm BH}}{M_\odot}
\right)
{\rm s}.
\end{equation}

For Cygnus X-1, with
\[
M_{\rm BH}\approx14.8M_\odot
\]
\citep{Orosz2011},
this yields
\[
\Delta t_{\rm phys}
=
5.40\times10^{-7}\times14.8\ {\rm s}
\approx
7.99\times10^{-6}\ {\rm s}
\approx
8\ \mu{\rm s}.
\]

For M87$^\ast$, with
\[
M_{\rm BH}\approx6.5\times10^9M_\odot
\]
\citep{EHT2019_VI},
the predicted asymmetry becomes
\[
\Delta t_{\rm phys}
=
5.40\times10^{-7}\times 6.5\times10^9\ {\rm s}
\approx
3.51\times10^3\ {\rm s}
\approx
58.5\ {\rm min}
\approx
1\ {\rm hr}.
\]

These examples illustrate how the same geometrized asymmetry naturally produces vastly different physical timing scales across stellar-mass and supermassive black holes through the linear dependence on black hole mass.

\section{Supplementary Figures}
\label{sec:appendixC}

This appendix presents additional delay maps, asymmetry visualizations, and parameter-space comparisons that supplement the primary figures discussed in the main text.

The supplementary figures include:
\begin{itemize}
    \item additional Kerr delay maps at intermediate spin values,
    \item prograde and retrograde asymmetry decompositions,
    \item normalized timing-asymmetry comparisons,
    \item extended inclination-dependent trends, and
    \item supplementary cross-sectional visualizations of the delay structure.
\end{itemize}

These supplementary figures further demonstrate the robustness and consistency of the directional timing asymmetry across the explored Kerr parameter space.
\end{document}